\documentstyle[12pt]{article}
\begin{document}
\begin{center}
{\bf THE BOOSTS IN THE NONCOMMUTATIVE SPECIAL RELATIVITY}
\vskip2truecm
M. Lagraa\\
\vskip1truecm
Laboratoire de Physique Th\'eorique\\
Universit\'e d'Oran Es-S\'enia, 31100, Alg\'erie\\
and\\
International Centre For Theoretical Physics, Trieste, Italy.\\
\vskip1truecm
\end{center}
{\bf Abstract :}
From the quantum analog of the Iwasawa decomposition of $SL(2,C)$ group and
the correspondence between quantum $SL(2,C)$ and Lorentz groups we deduce
the different properties of the Hopf algebra representing the boost of
particles in noncommutative special relativity. The representation of the
boost in the Hilbert space states is investigated and the addition rules of
the velocities are established from the coaction. The q-deformed
Clebsch-Gordon coefficients descibing the transformed states of the evolution
of particles in noncommutative special relativity are introduced and their
explicit calculation are given.
\newpage
\section{Introduction}
The Lorentz group play a fundamental role in special relativity. It gives the
lifetime dilatation of unstable particles in terms of their velocities and
relativistic formulas of the energy-momentum four-vector in terms of the mass
and the velocities.\\
The above considerations make especially interesting the study of the
noncommutative special relativity in the frame of the quantum Minkowski
space-time and its transformations under the quantum Lorentz group to derive
measurable observables describing the evolution of particles in noncommutative
space-time.\\
In the past few years, attention has been paid to formulate the particle
evolution in quantum Minkowski space times through the construction of the
$q$-analog of the relativistic plane waves [1] or the Hilbert space
representation of the $q$-deformed Minkowski space-time [2].\\
Despite all the theoretical interests, the relevance of the quantum Minkowski
space-time and its transformations under the quantum Lorentz group to derive
measurable observable effects in particle physics has not been discused very
much.\\
In Ref.[3] the evolution of particles in noncommutative Minkowski space-time
has been analysed. From the transformation of particle coordinates at
rest, the quantum analog of the lifetime dilatation of unstable particles and
the relativistic formulas of the energy-momentum four vector in terms of the
mass and the velocity have been established. The mean results of this work
concern the principle of the causality in the noncommutative special
relativity and the quantization of the moving particle lifetime which is
deduced from the discret spectrum of the velocity operators. In addition, it
is shown that for a particle moving in the noncommutative Minkowski space-time,
only the length of the velocity and its projection on the quantized direction
can be measured exactly.\\
In this paper we investigated further the tranformations of the Hilbert space
states describing particles moving in the noncommutative space-time to show
how the addition rule of the velocities can be deduced from the coaction of
the boost generators and how the quantum analog of Glebsch-Gordon coefficients
of the transformed states can be calculated.\\
This paper is organized as follows: In section 2, we present the quantum
analog of the Iwasawa decomposition of the $SL(2,C)$ group [4]. From this
decomposition and the correpondence between the quantum $SL(2,C)$ and Lorentz
groups [5], we extract the quantum boost generators and their commutation
relations. In section 3 we give a representation of these quantum
generators in the Hilbert space states. From the state transformation
principle and the coaction on the boost generators we establish the addition
rule of the velocities in the noncommutative special relativity and calculate
the quantum analog of the Clebsch-Gordon coefficients of the transformed
states.
\section{The quantum boost generators}
before we consider the Hopf algebra representing the quantum boost of
particles in the noncommutative Minkowski space-time let us briefly recall
the correspondence between the generators $\Lambda_{N}^{~M}~(N,~M=~0,~1,~2,~3)$
of the quantum Lorentz group and those of the quantum $SL(2,C)$ group generated
by $M_{\alpha}^{~\beta}$, $(\alpha, \beta=1,2)$ and
$(M_{\alpha}^{~\beta})^{\star}=M_{\dot{\alpha}}^{~\dot{\beta}}$ [5].\\
$M_{\alpha}^{~\beta}=\left(\begin{array}{cc}
\alpha& \beta\\ \gamma& \delta 
\end{array}
\right)$ corresponds to the representation of classical $SL(2,C)$ group
acting on space of undotted spinors and
$M_{\dot{\alpha}}^{~\dot{\beta}}$ corresponds to the classical $SL(2,C)$
group acting on space of dotted spinors.\\
The unimodularity of $M_{\alpha}^{~\beta}$ is expressed by
$\varepsilon_{\alpha\beta}M_{\gamma}^{~\alpha} M_{\delta}^{~\beta}=
\varepsilon_{\gamma\delta}I_{\cal A}$,
$\varepsilon^{\gamma\delta}M_{\gamma}^{~\alpha} M_{\delta}^{~\beta}=
\varepsilon^{\alpha\beta}I_{\cal A}$, and
$\varepsilon_{\dot{\alpha}\dot{\beta}}M_{\dot{\gamma}}^{~\dot{\alpha}}
M_{\dot{\delta}}^{~\dot{\beta}}=
\varepsilon_{\dot{\gamma}\dot{\delta}}I_{\cal A}$,
$\varepsilon^{\dot{\gamma}\dot{\delta}}M_{\dot{\gamma}}^{~\dot{\alpha}}
M_{\dot{\delta}}^{~\dot{\beta}}=
\varepsilon^{\dot{\alpha}\dot{\beta}}I_{\cal A}$ where $I_{\cal A}$ is the
unity of the $\star$ algebra ${\cal A}$ generated by $M_{\alpha}^{~\beta}$ and the
spinor metric $\varepsilon_{\alpha\beta}$ and its inverse
$\varepsilon^{\alpha\beta}~(\varepsilon_{\alpha\delta}
\varepsilon^{\delta\beta}=\delta_{\alpha}^{\beta}=
\varepsilon^{\beta\delta}\varepsilon_{\delta\alpha})$ satisfy
$(\varepsilon_{\alpha\beta})^{\star} =\varepsilon_{\dot{\beta}\dot{\alpha}}$
and $(\varepsilon^{\alpha\beta})^{\star}=
\varepsilon^{\dot{\beta}\dot{\alpha}}$. If we consider the case where the
quantum $SL(2,C)$ group admits the quantum $SU(2)$ group as a subgroup, the
spinor metric $\varepsilon_{\alpha\beta}$ must satisfy additional condition
$\varepsilon_{\alpha\beta}=-\varepsilon^{\dot{\beta}\dot{\alpha}}$ and
$\varepsilon^{\alpha\beta}=-\varepsilon_{\dot{\beta}\dot{\alpha}}$ required by
the compatibility of the unitarity,
$M_{(c)\dot{\alpha}}^{~~~~\dot{\beta}}=S(M_{(c)\beta}^{~~~~\alpha})=
\varepsilon_{\beta\rho}M_{(c)\sigma}^{~~~~\rho}\varepsilon^{\sigma\alpha}$, with
the modularity conditions of the quantum $SU(2)$ group generators. In this
case the commutation rules of the quantum $SU(2)$ subgroup are given by those
of $SL(2,C)$ group where we impose the unitarity condition. The commutation
rules are given by
\begin{eqnarray}
M_{\alpha}^{\rho}M_{\beta}^{\sigma}
R^{\pm\gamma\delta}_{~~\rho\sigma}=R^{\pm\rho\sigma}_{~~\alpha\beta}
M_{\rho}^{\delta}M_{\sigma}^{\delta},~~~and~~
M_{\dot{\alpha}}^{\dot{\rho}}M_{\dot{\beta}}^{\dot{\sigma}}
R^{\pm\dot{\gamma}\dot{\delta}}_{~~\dot{\rho}\dot{\sigma}}=
R^{\pm\dot{\rho}\dot{\sigma}}_{~~\dot{\alpha}\dot{\beta}}
M_{\dot{\rho}}^{\dot{\delta}}M_{\dot{\sigma}}^{\dot{\delta}}.
\end{eqnarray}
where the $R$-matrices are given by $R^{\pm\delta\beta}_{~\alpha\gamma}=
\delta^{\delta}_{\alpha}\delta^{\beta}_{\gamma} + q^{\pm 1}
\varepsilon^{\delta\beta}\varepsilon_{\alpha\gamma}~(
R^{\pm\dot{\delta}\dot{\beta}}_{~\dot{\alpha}\dot{\gamma}}=
\delta^{\dot{\delta}}_{\dot{\alpha}}\delta^{\dot{\beta}}_{\dot{\gamma}} +
q^{\pm 1}\varepsilon^{\dot{\delta}\dot{\beta}}
\varepsilon_{\dot{\alpha}\dot{\gamma}})$ and the spinor metric is of the form
$\varepsilon_{\alpha\beta}=\left(\begin{array}{cc}
0& -q^{-\frac{1}{2}}\\q^{\frac{1}{2}}&0 
\end{array}
\right)$ where $q\not=0$ is a real deformation parameter. The $R^{\pm}$-matrices
satisfy $R^{\pm\delta\beta}_{~\alpha\gamma}R^{\mp\rho\sigma}_{~\delta\beta}=
\delta^{\rho}_{\alpha}\delta^{\sigma}_{\gamma}~(
R^{\pm\dot{\delta}\dot{\beta}}_{~\dot{\alpha}\dot{\gamma}}
R^{\mp\dot{\rho}\dot{\sigma}}_{~\dot{\delta}\dot{\beta}}=
\delta^{\dot{\rho}}_{\dot{\alpha}}\delta^{\dot{\sigma}}_{\dot{\gamma}})$, the
Hecke conditions $(R^{\pm}+q^{\pm 2})(R^{\pm}-1)=0$ and the Yang-Baxter
equations. An explicit calculation gives from (1) the following commutation
rules of the quantum $SL(2,C)$ group as: 
\begin{eqnarray}
\alpha\beta=q\beta\alpha,~~~\alpha\gamma=q\gamma\alpha,
~~~\alpha\delta-q\gamma\beta=1,\nonumber\\
\gamma\delta=q\delta\gamma,~~~\gamma\beta=\beta\gamma,~~~
\beta\delta=q\delta\beta,~~~
\delta\alpha-q^{-1}\beta\gamma=1.
\end{eqnarray}\\
From the unitarity condition, the quantum $SU(2)$ subgroup reads
$M_{(c)\alpha}^{~~~~\beta}=
\left(\begin{array}{cc}
\alpha_{c}& -q\gamma_{c}^{\star}\\ \gamma_{c}& \alpha_{c}^{\star}, 
\end{array}
\right)$ and from (2) it follows that
\begin{eqnarray}
\alpha_{c}\alpha_{c}^{\star}+q^{2}\gamma_{c}\gamma_{c}^{\star}=1,~~~
\alpha_{c}^{\star}\alpha_{c}+\gamma_{c}\gamma_{c}^{\star}=1,~~~
\gamma_{c}\gamma_{c}^{\star}=\gamma_{c}^{\star}\gamma_{c},~~~
\alpha_{c}\gamma_{c}^{\star}=q\gamma_{c}^{\star}\alpha_{c},~~~
\alpha_{c}\gamma_{c}=q\gamma_{c}\alpha_{c}.
\end{eqnarray}
It is shown in [5] that the generators $\Lambda_{N}^{~M}$ of quantum Lorentz
group may be written in terms of those of quantum $SL(2,C)$ group as
\begin{eqnarray}
\Lambda_{N}^{~M} =\frac{1}{Q}\varepsilon_{\dot{\gamma}\dot{\delta}}
\overline{\sigma}_{N}^{~\dot{\delta}\alpha}M_{\alpha}^{~\sigma}
\sigma^{M}_{~\sigma\dot{\rho}}M_{\dot{\beta}}^{~\dot{\rho}}
\varepsilon^{\dot{\gamma}\dot{\beta}}.
\end{eqnarray}
They are real, $(\Lambda_{N}^{~M})^{\star}=\Lambda_{N}^{~M}$, and generate a
Hopf algebra $\cal L$ endowed with a coaction $\Delta$, a counit
$\varepsilon$ and an antipode $S$ acting as $\Delta(\Lambda_{N}^{~M}) =
\Lambda_{N}^{~K}\otimes \Lambda_{K}^{~M}$, $\varepsilon(\Lambda_{N}^{~M})=
\delta_{N}^{M}$ and $S(\Lambda_{N}^{~M})=
G_{\pm NK}\Lambda_{L}^{~K}G_{\pm}^{LM}$ respectively. $G_{\pm}^{NM}$ is an
invertible and hermitian quantum metric. It may be expressed in
terms of the four matrices $\sigma^{N}_{\alpha\dot{\beta}}~(N=0,1,2,3)$, where
$\sigma^{n}_{\alpha\dot{\beta}}~(n=1,2,3)$ are the usual Pauli matrices and
$\sigma^{0}_{\alpha\dot{\beta}}$ is the identity matrix, as:
\begin{eqnarray*}
G_{\pm}^{~IJ} =\frac{1}{Q}Tr(\sigma^{I} \overline{\sigma}_{\pm}^{J}) =
\frac{1}{Q}\varepsilon^{\alpha \nu}\sigma^{I}_{~\alpha\dot{\beta}}
\overline{\sigma}_{\pm}^{J\dot{\beta}\gamma}\varepsilon_{\gamma\nu}=
\frac{1}{Q}Tr(\overline{\sigma}_{\pm}^{I} \sigma^{J})=
\frac{1}{Q} \varepsilon_{\dot{\nu}\dot{\gamma}}
\overline{\sigma}_{\pm}^{I\dot{\gamma}\alpha}
\sigma^{J}_{~\alpha\dot{\beta}} \varepsilon^{\dot{\nu}\dot{\beta}}
\end{eqnarray*}
where
$\overline{\sigma}_{\pm}^{I\dot{\alpha} \beta} = \varepsilon^{\dot{\alpha}
\dot{\lambda}}R^{\mp \sigma \dot{\rho}}_{~~\dot{\lambda}\nu}
\varepsilon^{\nu \beta}\sigma^{I}_{~\sigma \dot{\rho}}$. The undotted and
dotted spinorial indices are raised and lowered as
$\sigma^{I\alpha}_{~~\dot{\beta}} = \sigma^{I}_{~\rho\dot{\beta}}
\varepsilon^{\rho\alpha}$ and $\sigma^{I~\dot{\beta}}_{~\alpha}=
\varepsilon^{\dot{\beta}\dot{\rho}}\sigma^{I}_{\alpha\dot{\rho}}$ and the
inverse of the metric may be written under the form
$G_{\pm IJ}=\frac{1}{Q}Tr(\overline{\sigma}_{J}\sigma_{\pm I})=
\frac{1}{Q} \varepsilon_{\dot{\nu}\dot{\gamma}}
\overline{\sigma}_{J}^{~\dot{\gamma}\alpha}
\sigma_{\pm I\alpha\dot{\beta}} \varepsilon^{\dot{\nu}\dot{\beta}}$ where
$\sigma_{\pm I\alpha \dot{\beta}} = G_{\pm IJ}\sigma^{J}_{\alpha\dot{\beta}}$.
The form of the antipode of $\Lambda_{N}^{~M}$ implies the orthogonality
condition on the generators of quantum Lorentz group as:
\begin{eqnarray}
G_{\pm NM}\Lambda_{L}^{~N}\Lambda_{K}^{~M} = G_{\pm LK}I_{\cal L}~~and~~
G_{\pm}^{~LK}\Lambda_{L}^{~N}\Lambda_{K}^{~M} = G_{\pm}^{~NM}I_{\cal L}
\end{eqnarray}
where $I_{{\cal L}}$ is the unity of the Hopf algebra $\cal L$.\\
In this framwork, it is also shown that there exist two copies of
the quantum Minkowski space-times ${\cal M}_{\pm}$ equipped with metric
$G^{\pm IK}$ and real coordinates $X_{\pm I}$ which transform under the left
coaction
$\Delta_{L}:{\cal M}_{\pm} \rightarrow {\cal L}\otimes {\cal M}_{\pm}$ as:
\begin{eqnarray}
\Delta_{L}(X_{\pm I})=\Lambda_{I}^{~K}\otimes X_{\pm K}.
\end{eqnarray}
The coordinates $X_{+I}$ transform under the quantum lorentz whose generators
$\Lambda_{N}^{~M}$ are subject to commutation rules controlled by the
${\cal R}_{~PQ}^{+NM}$,
$\Lambda_{L}^{~P}\Lambda_{K}^{~Q}{\cal R}_{~PQ}^{+NM}=
{\cal R}_{~LK}^{+PQ}\Lambda_{P}^{~N}\Lambda_{Q}^{~M}$,  and the coordinates
$X_{-I}$ transform under the quantum lorentz whose generators
$\Lambda_{N}^{~M}$ are subject to commutation rules of the form
$\Lambda_{L}^{~P}\Lambda_{K}^{~Q}{\cal R}_{~PQ}^{-NM}=
{\cal R}_{~LK}^{-PQ}\Lambda_{P}^{~N}\Lambda_{Q}^{~M}$ where the
${\cal R}$-matrices of the Lorentz group are constructed out of those of
$SL(2,C)$ group as:
\begin{eqnarray*}
{\cal R}_{~LK}^{\pm NM} =\frac{1}{Q^{2}}R^{\mp\alpha\kappa}_{~\tau\sigma}
R^{\mp\mu\beta}_{~\kappa\nu}R^{\pm\rho\epsilon}_{~\delta\alpha}
R^{\pm\lambda\gamma}_{~\epsilon\mu}\sigma^{M~\dot{\beta}}_{~\gamma}
\sigma^{N~\dot{\lambda}}_{~\rho} \overline{\sigma}_{L\dot{\tau}}^{~~\delta}
\overline{\sigma}_{K\dot{\nu}}^{~~\sigma}.
\end{eqnarray*}
These ${\cal R}$-matrices lead to the symmetrization of the Minkowski metric
in the quantum sens as:
\begin{eqnarray*}
{\cal R}^{\pm NM}_{~KL}G_{\pm}^{KL} = G_{\pm}^{NM}~~,~~
{\cal R}^{\pm NM}_{~KL}G_{\pm NM} = G_{\pm KL}
\end{eqnarray*}
and satisfy the Yang-Baxter equations and the cubic Hecke conditions
\begin{eqnarray*}
({\cal R}^{\pm} + q^{\pm2})({\cal R}^{\pm} + q^{\mp2})({\cal R}^{\pm} - 1)=0.
\end{eqnarray*}
In the following we shall consider the right invariant basis $X_{I}=X_{+I}$ as a
quantum coordinate system of the Minkowski space-time ${\cal M}={\cal M}_{+}$
equipped with the metric $G^{IJ}=G_{+}^{~IJ}$. $X_{0}$ represents the time
operator and $X_{i}~(i=1,2,3)$ represent the space coordinate operators.\\
With this choice, the commutation rules between the undotted and dotted
generators of the quantum $SL(2,C)$ group must necessarely be
controlled by the $R^{-}$-matrix,
$M_{\alpha}^{\gamma}M_{\dot{\sigma}}^{\dot{\delta}}
R^{-\beta\delta}_{~\rho\gamma}=R^{-\gamma\sigma}_{~\delta\alpha}
M_{\dot{\delta}}^{\dot{\rho}}M_{\gamma}^{\beta}$ which gives
explicitly
\begin{eqnarray}
\alpha\alpha^{\star}=\alpha^{\star}\alpha-(1-q^{-2})\beta\beta^{\star}&,&~~
\alpha\gamma^{\star}=q^{-1}\gamma^{\star}\alpha -(1-q^{-2})\beta\delta^{\star},
\nonumber\\
\gamma\gamma^{\star}=\gamma^{\star}\gamma+(1-q^{-2})(\alpha\alpha^{\star}-
\delta\delta^{\star})&,&~~
\gamma\delta^{\star}=q\delta^{\star}\gamma +q(1-q^{-2})\beta^{\star}\alpha
\nonumber\\
\beta\beta^{\star}=\beta^{\star}\beta,~~~
\beta\delta^{\star}=q^{-1}\delta^{\star}\beta&,&~~\delta\delta^{\star}=
\delta^{\star}\delta+(1-q^{-2})\beta^{\star}\beta,\nonumber\\
\alpha\beta^{\star}=q\beta^{\star}\alpha,~~~ \alpha\delta^{\star}=
\delta^{\star}\alpha&,&~~\gamma\beta^{\star}=\beta^{\star}\gamma
\end{eqnarray}
Note that if we have considered the coordinates $X_{I}=X_{-I}$ wich
correspond to the metric $G^{NM}=G^{-NM}$ and the ${\cal R}$-matrix
$R^{NM}_{KL}=R^{-NM}_{~KL}$ the commutation rules between the undotted and
dotted generators of the quantum $SL(2,C)$ group must necessarely be
controlled by the $R^{+}$-matrix.\\
To make an explicit calculation of the different commutation rules of the
generators of the quantum Lorentz group, we take the following apropriate choice
of Pauli hermitian matrices
\begin{eqnarray*}
\sigma^{0}_{\alpha \dot{\beta}} =\left(\begin{array}{cc}
1 & 0\\
0 & 1
\end{array}
\right)~~,~~
\sigma^{1}_{\alpha \dot{\beta}} =\left(\begin{array}{cc}
0 & 1\\1 & 0
\end{array}
\right)~~,~~
\sigma^{2}_{\alpha \dot{\beta}} =\left(\begin{array}{cc}
0 & -i\\
i &  0
\end{array}
\right)~~,~~
\sigma^{3}_{\alpha \dot{\beta}} =\left(\begin{array}{cc}
q & 0\\
0 &-q^{-1}
\end{array}
\right).
\end{eqnarray*}
The advantage of this choice arises from the fact that:\\
1)- The quantum metric $G^{LK}$ is of the form of two independent blocks,
one for the time index and the others for space component indeces
$(k=1,2,3)$. The nonvanishing elements of the metrics are;
$G^{00}=-q^{-\frac{3}{2}}$, $G^{11}=G^{22}=G^{33}=q^{\frac{1}{2}}$,
$G^{12}=-G^{21}=-iq^{\frac{1}{2}}\frac{q-q^{-1}}{Q}$ and the non vanishing
elements of its inverse are $G_{00}=-q^{\frac{3}{2}}$, $G_{11}=
G_{22}=q^{-\frac{1}{2}}\frac{Q^{2}}{4}$, $G_{33}=q^{-\frac{1}{2}}$ and
$G_{12}=-G_{21}=iq^{-\frac{1}{2}}\frac{(q-q^{-1})Q}{4}$. In the classical
limit $q=1$, this metric reduces to the classical Minkowski metric with
signature $(-,+,+,+)$.\\
2)- $\overline{\sigma}_{0}^{~\dot{\alpha}\beta} =
-\sigma^{0}_{\alpha \dot{\beta}}=-\delta_{\alpha}^{\beta}$,
$\overline{\sigma}_{N\dot{\alpha}\alpha}=\overline{\sigma}_{N\dot{1}1} +
\overline{\sigma}_{N\dot{2}2}=-(q+q^{-1})\delta_{N}^{0}=-Q\delta_{N}^{0} $
and $\sigma^{N\alpha\dot{\alpha}}=Q\delta_{0}^{N}$. These properties make
explicit the restriction of the quantum Lorentz group to the quantum subgroup
of the three dimensional space rotations by restricting the quantum $SL(2,C)$
group generators to those of the $SU(2)$ group. In fact when we restrict the
generators of the quantum $SL(2,C)$ group to those of the $SU(2)$ by imposing
unitarity condition in (4), we get [3]
\begin{eqnarray}
\Lambda_{N}^{~0} =\frac{1}{Q}
\overline{\sigma}_{N\dot{\gamma}}^{~~~\alpha}M_{\alpha}^{~\sigma}
\sigma^{0}_{~\sigma\dot{\rho}}S(M_{\rho}^{~\beta})
\varepsilon^{\dot{\gamma}\dot{\beta}} =\frac{1}{Q}
\overline{\sigma}_{N\dot{\gamma}}^{~~~\alpha}
\varepsilon^{\dot{\gamma}\dot{\alpha}}
=-\frac{1}{Q}\overline{\sigma}_{N\dot{\gamma}\gamma}= \delta_{N}^{0}
\end{eqnarray}
and
\begin{eqnarray}
\Lambda_{0}^{~M} =\frac{1}{Q}\varepsilon_{\dot{\gamma}\dot{\delta}}
\overline{\sigma}_{0}^{~\dot{\delta}\alpha}M_{\alpha}^{~\sigma}
\sigma^{M}_{~\sigma\dot{\rho}}S(M_{\rho}^{~\beta})
\varepsilon^{\dot{\gamma}\dot{\beta}} &=& -\frac{1}{Q}\varepsilon^{\alpha\gamma}
M_{\alpha}^{~\sigma}\sigma^{M}_{~\sigma\dot{\rho}}\varepsilon_{\rho\delta}
M_{\gamma}^{~\delta}=\nonumber\\
\frac{1}{Q} \varepsilon^{\dot{\delta}\dot{\rho}}
\sigma^{M}_{~\sigma\dot{\rho}}\varepsilon^{\sigma\delta} &=& \frac{1}{Q}
\sigma^{M\delta\dot{\delta}}=\delta_{0}^{M}
\end{eqnarray}
which lead us to the restriction of the Minkowski space-time transformations
to the orthogonal transformations of the three dimensional space $R_{3}$
equipped with the coordinate system $X_{i}$, ($i=1,2,3$). These
transformations leave invariant the time coordinate $X_{0}$. In fact, as a
consequence of (8) and (9) we have
\begin{eqnarray}
\Delta_{L}(X_{0}) = \overline{\Lambda}_{0}^{~0}\otimes X_{0}
= I\otimes X_{0}\nonumber\\
\Delta_{L}(X_{i}) = \overline{\Lambda}_{i}^{~j}\otimes X_{j}
\end{eqnarray}
where $\overline{\Lambda}_{i}^{~j}=
\frac{1}{Q}
\overline{\sigma}_{i\dot{\gamma}}^{~~\alpha}M_{(c)\alpha}^{~~~~\sigma}
\sigma^{j}_{~\sigma\dot{\rho}}M_{(c)\dot{\beta}}^{~~~~\dot{\rho}}
\varepsilon^{\dot{\gamma}\dot{\beta}}=\frac{1}{Q}
\overline{\sigma}_{i\dot{\gamma}}^{~~\alpha}M_{(c)\alpha}^{~~~~\sigma}
\sigma^{j}_{~\sigma\dot{\rho}}S(M_{(c)\rho}^{~~~~\beta})
\varepsilon^{\dot{\gamma}\dot{\beta}}$ generate a Hopf subalgebra
${\cal SO}_{q}(3)$ of $\cal L$ whose the axiomatic structure is derived from
those of $\cal L$ as $\Delta(\overline{\Lambda}_{i}^{~j}) =
\overline{\Lambda}_{i}^{~k}\otimes \overline{\Lambda}_{k}^{~j}$,
$\varepsilon(\overline{\Lambda}_{i}^{~j})= \delta_{i}^{~j}$ and
$S(\overline{\Lambda}_{i}^{~j}) = G_{iK}
\overline{\Lambda}_{L}^{~K}G^{~Lj} = G_{ik}
\overline{\Lambda}_{l}^{~k}G^{~lj}$ where
$G^{~ij}$ is the restriction of the quantum Minkowskian metric $G^{~IJ}$. It
is the Euclidian metric of the quantum space $R_{3}$ satisfying
$G^{~ik}G_{kj} = \delta_{j}^{i} =G_{jk}G^{~ki}$.
The form of the antipode of $\overline{\Lambda}_{i}^{~j}$ implies
the orthogonality properties
\begin{eqnarray*}
G^{~ij} \overline{\Lambda}_{i}^{~l}\overline{\Lambda}_{j}^{~k}
= G^{~lk} &~and&
G_{lk} \overline{\Lambda}_{i}^{~l} \overline{\Lambda}_{i}^{~k}= G_{ij}.
\end{eqnarray*}
Therefore, $\overline{\Lambda}_{i}^{~j}=\frac{1}{Q}
\overline{\sigma}_{i\dot{\gamma}}^{~~\alpha}M_{(c)\alpha}^{~~~~\sigma}
\sigma^{j}_{~\sigma\dot{\rho}}S(M_{(c)\rho}^{~~~~\beta})
\varepsilon^{\dot{\gamma}\dot{\beta}}$ establishes a correspondence between
$SU_{q}(2)$ and $SO_{q}(3)$ group. In the three dimensional space spanned by
the basis $Z$, $\overline{Z}$ and $X_{3}$, the generators
$\overline{\Lambda}_{i}^{~j}=\frac{1}{Q}
\overline{\sigma}_{i\dot{\gamma}}^{~~\alpha}M_{(c)\alpha}^{~~~~\sigma}
\sigma^{j}_{~\sigma\dot{\rho}}S(M_{(c)\rho}^{~~~~\beta})
\varepsilon^{\dot{\gamma}\dot{\beta}}$ give the irreducible three-dimensional
representation of $SU_{q}(2)$ considered in [6] as
\begin{eqnarray}
(d_{1,i}^{~~~j})=
\left(\begin{array}{clcr}
-2q\gamma_{c}\gamma_{c} & 2\alpha_{c}^{\star}\alpha_{c}^{\star}&
Q\gamma_{c}\alpha_{c}^{\star}\\
2\alpha_{c}\alpha_{c} & -2q\gamma_{c}^{\star}\gamma_{c}^{\star} &
Q\alpha_{c}\gamma_{c}^{\star}\\
-2\alpha_{c}\gamma_{c}& -2\gamma_{c}^{\star}\alpha_{c}^{\star} &
1-qQ\gamma_{c}\gamma_{c}^{\star}
\end{array}
\right)\in M_{3}\otimes C(SU_{q}(2))
\end{eqnarray}
where the indices $i,j$ run over $z=1+i2$ and $\overline{z}=1-i2$.\\
It is shown in [4] that the quantum $SL(2,C)$ group admits an unique Iwasawa
decomposition of the form $M_{\alpha}^{~\beta} = M_{(c)\alpha}^{~~~~\rho}
M_{(d)\rho}^{~~~~\beta}$ where $M_{(c)}$ is a quantum $SU(2)$-matrix and
$M_{(d)}$ is a quantum-matrix the left-lower-corner element equal to zero. The
matrix elements of $M_{(c)}$ doubly commute with matrix elements of $M_{(d)}$
(two operators $a$ and $b$ doubly commute if $ab=ba$ and
$a b^{\star}=b^{\star}a$).\\
With our choice of the commutation rules (7) between
the undotted and dotted generators of the quantum $SL(2,C)$ group, in order
to have nontrivial commutation relations, $M_{(d)}$ must have the
right-upper-corner element equal to zero
\begin{eqnarray}
M_{(d)\alpha}^{~~~~\beta}=\left(\begin{array}{cc}
a& 0\\
n& a^{-1}
\end{array}
\right).
\end{eqnarray}
From (12) and (7), it follows that 
\begin{eqnarray}
an=qna,~~~aa^{\star}=a^{\star}a,~~~an^{\star}=q^{-1}n^{\star}a,\\
nn^{\star}=n^{\star}n + (1-q^{-2})(aa^{\star} -(aa^{\star})^{-1}).
\end{eqnarray}
As $M_{(d)}$ is a subgroup of the quantum $SL(2,C)$ with coaction, counity and
antipode acting in the following way: $\Delta(a) =a \otimes a$, and
$\Delta(n) =n \otimes a + a^{-1} \otimes n$, $\varepsilon (a)=1$,
$\varepsilon (n)=0$, $S(a) = a^{-1}$, and $S(n) =-qn$. The Iwasawa
decomposition and the correspondence between quantum $SL(2,C)$ and Lorentz
group (4) permits us to extract out of the latter the $SO_{q}(3)$ subgroup
left by restriction to $M_{(d)}$ with the quantum boost.\\
More precisely by replacing the matrix elements of $M$ by those of $M_{(d)}$
into the relation (4), we get the following generators of the quantum boost
as:
\begin{eqnarray}
\Lambda_{0}^{~0} =\frac{1}{Q}(q^{-1}aa^{\star} + q(aa^{\star})^{-1} +
q^{-1}nn^{\star})&,&~~\Lambda_{3}^{~3} = \frac{1}{Q}(qaa^{\star} +
q^{-1}(aa^{\star})^{-1} - qnn^{\star}),\nonumber\\
\Lambda_{3}^{~0} = \frac{1}{Q}(aa^{\star} - (aa^{\star})^{-1} - nn^{\star})&,&~~
\Lambda_{0}^{~3} =\frac{1}{Q}(aa^{\star} -(aa^{\star})^{-1} +q^{2}nn^{\star})
\nonumber\\
\Lambda_{z}^{~0} = na^{\star},~~\Lambda_{z}^{~3} = qna^{\star}&,&~~
\Lambda_{0}^{~z} = \frac{2q}{Q}n(a^{\star})^{-1},~~\Lambda_{3}^{~z} =
\frac{-2}{Q}n(a^{\star})^{-1}\nonumber\\
\Lambda_{z}^{~\overline{z}} = 2a^{-1}a^{\star}&,&~~\Lambda_{z}^{~z} = 0.
\end{eqnarray}
The remaining generators are obtained by complex conjugation. Note that the
commutation relations (13-14) permit us to take $a$
real, $a^{\star}=a$, then $\Lambda_{z}^{~\overline{z}}=\Lambda_{\overline{z}}
^{~z}=2$.\\
From the commutation relations (13-14), we see that $[a^{-1}a^{\star},n]=
[a^{-1}a^{\star},a]=0$ which implies that $\Lambda_{z}^{~\overline{z}}$ is
central, i.e. $[\Lambda_{z}^{~\overline{z}},\Lambda_{N}^{~M}]=0$. From
$[a,nn^{\star}]=[a^{\star},nn^{\star}]=0$ we deduce that
$[\Lambda_{3}^{~3},\Lambda_{0}^{~3}] = [\Lambda_{3}^{~3},\Lambda_{3}^{~0}] =
[\Lambda_{0}^{~3},\Lambda_{3}^{~0}]=0$. A straighforward computation shows
that $\Lambda_{0}^{~0}$ is central, $[\Lambda_{0}^{~0},\Lambda_{N}^{~M}]=0$,
and
\begin{eqnarray}
\Lambda_{3}^{~0}\Lambda_{z}^{~0}-q^{2}\Lambda_{z}^{~0}\Lambda_{3}^{~0}&=&
(q-q^{-1})\Lambda_{0}^{~0}\Lambda_{z}^{~0},\\
\Lambda_{z}^{~0}\Lambda_{\overline{z}}^{~0}-\Lambda_{\overline{z}}^{~0}
\Lambda_{z}^{~0}&=&
(q-q^{-1})Q\Lambda_{3}^{~0}(\Lambda_{3}^{~0}+q^{-1}\Lambda_{0}^{~0}),\\
\Lambda_{3}^{~0}\Lambda_{0}^{~z}-q^{2}\Lambda_{0}^{~z}\Lambda_{3}^{~0}&=&
(q-q^{-1})\Lambda_{0}^{~0}\Lambda_{0}^{~z},\\
\Lambda_{0}^{~z}\Lambda_{\overline{z}}^{~0}-q^{2}\Lambda_{\overline{z}}^{~0}
\Lambda_{0}^{~z}&=&
q^{2}(q-q^{-1})\Lambda_{3}^{~0}\Lambda_{\overline{z}}^{~z},\\
\Lambda_{0}^{~3}\Lambda_{0}^{~z}-q^{-2}\Lambda_{0}^{~z}\Lambda_{0}^{~3}&=&
(q-q^{-1})\Lambda_{0}^{~0}\Lambda_{0}^{~z},\\
\frac{Q}{4}(\Lambda_{0}^{~\overline{z}}\Lambda_{0}^{~z}-\Lambda_{0}^{~z}
\Lambda_{0}^{~\overline{z}})&=&
(q-q^{-1})\Lambda_{0}^{~3}(\Lambda_{0}^{~3}-q\Lambda_{0}^{~0}),\\
\Lambda_{0}^{~z}\Lambda_{z}^{~0}-q^{-2}\Lambda_{z}^{~0}\Lambda_{0}^{~z}&=&0\\
\Lambda_{3}^{~3}\Lambda_{z}^{~0}-q^{2}\Lambda_{z}^{~0}\Lambda_{3}^{~3}&=&
q^{-1}(q-q^{-1})\Lambda_{z}^{~0}
(\Lambda_{0}^{~3}+\frac{Q}{4}\Lambda_{z}^{~\overline{z}}),\\
\Lambda_{3}^{~3}\Lambda_{0}^{~z}-q^{-2}\Lambda_{0}^{~z}\Lambda_{3}^{~3}&=&
-q(q-q^{-1})\Lambda_{0}^{~z}
(\Lambda_{3}^{~0}-\frac{1}{Q}\Lambda_{\overline{z}}^{~z}).
\end{eqnarray}
The remaining commutation relations are obtained by substituting into (16-24)
the following relations
\begin{eqnarray}
\Lambda_{z}^{~0}=q^{-1}\Lambda_{z}^{~3},~~~
\Lambda_{0}^{~z}=-q\Lambda_{3}^{~z},~~~
\Lambda_{3}^{~3}+q^{-1}\Lambda_{0}^{~3}=
\Lambda_{0}^{~0}+q\Lambda_{3}^{~0}
\end{eqnarray}
deduced from the form of the boost generators (15). From these relations and
the commutation rules (16-24) we can also show that the orthogonality
condition $G^{NM}\Lambda_{N}^{~L}\Lambda_{M}^{~K}=G^{LK}$ and
$G_{LK}\Lambda_{N}^{~L}\Lambda_{M}^{~K}=G_{NM}$ are satisfied if
\begin{eqnarray}
\Lambda_{z}^{~0}=\frac{Q}{4}\Lambda_{z}^{~\overline{z}}\Lambda_{0}^{~z}
(q^{-1}\Lambda_{0}^{~0}+\Lambda_{3}^{~0}),~~~
\Lambda_{\overline{z}}^{~0}=q^{-2}\frac{Q}{4}\Lambda_{\overline{z}}^{~z}
\Lambda_{0}^{~\overline{z}}(q^{-1}\Lambda_{0}^{~0}+\Lambda_{3}^{~0}),~~~\\
\Lambda_{0}^{~z}=\frac{1}{Q}\Lambda_{\overline{z}}^{~z}\Lambda_{z}^{~0}
(q\Lambda_{0}^{~0}-\Lambda_{0}^{~3}),~~~
\Lambda_{0}^{~\overline{z}}=q^{2}\frac{1}{Q}\Lambda_{z}^{~\overline{z}}
\Lambda_{\overline{z}}^{~0}(q\Lambda_{0}^{~0}-\Lambda_{0}^{~3}).
\end{eqnarray}
In fact by substituting (15) and (25) into
$G^{NM}\Lambda_{N}^{~z}\Lambda_{M}^{~0}=0 =
-q^{-\frac{3}{2}}\Lambda_{0}^{~z}\Lambda_{0}^{~0}+
q^{\frac{1}{2}}(\frac{q\Lambda_{z}^{~z}
\Lambda_{\overline{z}}^{~0}+q^{-1}\Lambda_{\overline{z}}^{~z}
\Lambda_{z}^{~0}}{Q}+\Lambda_{3}^{~z}\Lambda_{3}^{~0})$, we get
\begin{eqnarray*}
\frac{q^{-1}}{Q}\Lambda_{\overline{z}}^{~z}\Lambda_{z}^{~0}=
q^{-2}\Lambda_{0}^{~z}\Lambda_{0}^{~0} +
q^{-1}\Lambda_{0}^{~z}\Lambda_{3}^{~0}=
q^{-1}\Lambda_{0}^{~z}(q^{-1}\Lambda_{0}^{~0}+\Lambda_{3}^{~0})
\end{eqnarray*}
leading to the left relation of (26). The same procedure gives from
$G^{NM}\Lambda_{N}^{~\overline{z}}\Lambda_{M}^{~0}=0$ the right relation of
(26) and from $G_{NM}\Lambda_{z}^{~N}\Lambda_{0}^{~M}=0$ and
$G_{NM}\Lambda_{\overline{z}}^{~N}\Lambda_{0}^{~M}=0$ the relations
(27).\\
By substituting (27) into (26) and by using
$\Lambda_{z}^{~\overline{z}}\Lambda_{\overline{z}}^{~z}=4$ obtained from (15),
we get $(q\Lambda_{0}^{~0}-\Lambda_{0}^{~3})
(q^{-1}\Lambda_{0}^{~0}+\Lambda_{3}^{~0})=1$ leading to
\begin{eqnarray}
\Lambda_{3}^{~3}=q\Lambda_{3}^{~0}+
(\Lambda_{0}^{~0}+q\Lambda_{3}^{~0})^{-1},~~~
\Lambda_{0}^{~3}=q\Lambda_{0}^{~0}-
q(\Lambda_{0}^{~0}+q\Lambda_{3}^{~0})^{-1}.
\end{eqnarray}
Then all generators of the boost can be given in terms of $\Lambda_{0}^{~0}$,
$\Lambda_{z}^{~0}$, $\Lambda_{\overline{z}}^{~0}$ and $\Lambda_{3}^{~0}$. As
in [3], we set $\Lambda_{0}^{~0}=\gamma$ and $\Lambda_{i}^{~0}=\gamma V_{i}$
where $\gamma$ is a real c-number given by
\begin{eqnarray}
\gamma =(1-|v|_{q}^{2})^{-\frac{1}{2}},
\end{eqnarray}
$V_{i}$ are the components of the velocity operator and
$|v|_{q}^{2}=-\frac{G^{ij}}{G^{00}}V_{i}V_{j}$ is its lenght which is also a
c-number.
\section{\bf The addition rules of velocities in the noncommutative special
relativity}
In [3] the evolution of free particles in the coordinate system
$X_{N}$, $(N=0,1,2,3)$ of the Minkowski space-time are
described in terms of states $|{\cal P}\rangle=|t,x_{3},\tau^{2}\rangle$
belonging to the Hilbert space ${\cal H}_{\cal M}$. Here $t$, $x_{3}$ and
$\tau^{2}$ are the time, the coordinate $x_{3}$ and the proper-time
respectively. They are eingenvalues of the set of commuting operators which
are the time $X_{0}$, the component $X_{3}$ of the space coordinates and
the lenght of the four-vector $X_{N}$, $G^{NM}X_{N}X_{M}=-\tau^{2}=
-q^{-\frac{3}{2}}X_{0}^{2} + \frac{q^{\frac{3}{2}}}{Q}Z\overline{Z} +
\frac{q^{-\frac{1}{2}}}{Q}\overline{Z}Z + q^{\frac{1}{2}}X_{3}^{2}$, where
$Z=X_{1}+iX_{2}$ and $\overline{Z}=X_{1}-iX_{2}$. $\tau^{2}$ is real,
bi-invariant and central. Then $|{\cal P}\rangle$ is a common eingenstate of
$X_{0}$, $X_{3}$ and $\tau^{2}$
\begin{eqnarray}
X_{0}|{\cal P}\rangle=t|{\cal P}\rangle,~~~X_{3}|{\cal P}\rangle=
x_{3}|{\cal P}\rangle,~~~and~~~\tau^{2}|{\cal P}\rangle=
\tau^{2}|{\cal P}\rangle.
\end{eqnarray}
Since the coordinate system transforms under quantum Lorentz group with
tensorial product as $X'_{N}=\Lambda_{N}^{~M}\otimes X_{M}$ we have assumed in
[3] that the Hilbert state $|{\cal P}\rangle$ transforms into
$|{\cal P}'\rangle$ as
\begin{eqnarray}
|{\cal P}'\rangle =|sym_{q}\rangle\otimes |{\cal P}\rangle
\end{eqnarray}
where $|{\cal P}'\rangle$ describes the evolution of the particle in the
coordinate system $X'_{N}$. Note that the coordinates $X'_{N}$ fulfil the
same commutation relations as those of $X_{N}$, then
$|{\cal P}'\rangle=|t',x'_{3},\tau^{2}\rangle$ satisfy also
\begin{eqnarray}
X'_{0}|{\cal P}'\rangle=t'|{\cal P}'\rangle,~~~X'_{3}|{\cal P}'\rangle=
x'_{3}|{\cal P}'\rangle,~~~and~~~\tau^{2}|{\cal P}'\rangle=
\tau^{2}|{\cal P}'\rangle.
\end{eqnarray}
The state $|sym_{q}\rangle$ belongs to the Hilbert state
${\cal H}_{\cal L}$ where the quantum Lorentz generators act. It is a common
eingenstate of a set of commuting operators of (15). Since all the generators
of the Boost can be written in terms of $\Lambda_{N}^{~0}=\gamma V_{N}$, the
set of commuting operators of the Boost
are $\Lambda_{0}^{~0}=\gamma$ and $\Lambda_{3}^{~0}=\gamma V_{3}$ and
therefore, $|sym_{q}\rangle= |v_{3},\gamma\rangle$ is a common eingenstate of
$\gamma$ or the lenght of the velocity $-\frac{G^{ij}}{G^{00}}V_{i}V_{j}$ and
$V_{3}$ 
\begin{eqnarray}
\gamma|v_{3},\gamma\rangle=\gamma|v_{3},\gamma\rangle~~and~~
V_{3}|v_{3},\gamma\rangle=v_{3}|v_{3},\gamma\rangle.
\end{eqnarray}
The coordinates $X'_{N}$ act on the transformed state $|{\cal P}'\rangle$ as  
\begin{eqnarray}
X'_{0}|{\cal P}'\rangle = 
(\Lambda_{0}^{~0}\otimes X_{0})|{\cal P}'\rangle +
(\Lambda_{0}^{~k}\otimes X_{k})|{\cal P}'\rangle=
\Lambda_{0}^{~0}|v_{3}\gamma\rangle\otimes X_{0}|{\cal P}\rangle +
\Lambda_{0}^{~k}|v_{3}\gamma\rangle\otimes X_{k}|{\cal P}\rangle,\\
X'_{i}|{\cal P}'\rangle = 
(\Lambda_{i}^{~0}\otimes X_{0})|{\cal P}'\rangle +
(\Lambda_{i}^{~k}\otimes X_{k})|{\cal P}'\rangle=
\Lambda_{i}^{~0}|v_{3}\gamma\rangle \otimes X_{0}|{\cal P}\rangle +
\Lambda_{i}^{~k}|v_{3}\gamma\rangle \otimes X_{k}|{\cal P}\rangle.
\end{eqnarray}
In the case where we boost a particle at rest described by the state
$|{\cal P}_{0}\rangle=|t,0,\tau^{2}\rangle$ satisfying
$X_{0}|{\cal P}_{0}\rangle=t|{\cal P}_{0}\rangle$,
$X_{i}|{\cal P}_{0}\rangle=0 |{\cal P}_{0}\rangle$ and
$\tau^{2}|{\cal P}_{0}\rangle=\tau^{2}|{\cal P}_{0}\rangle$ it is shown in
[3] that this state is unique and transforms under the quantum Lorentz group
as:
\begin{eqnarray}
|{\cal P}\rangle=|t,x_{3},\tau^{2}\rangle
=|v_{3},\gamma\rangle\otimes|{\cal P}_{0}\rangle.
\end{eqnarray}
$x_{3}$ and $v_{3}$ are quantized and read
\begin{eqnarray}
x_{3}^{(l,m)}=q^{-1}(\frac{q^{2m}}{\gamma^{(l)}} - 1)t~~,~~
v_{3}^{(l,m)}= q^{-1}(\frac{q^{2m}}{\gamma^{(l)}} - 1)
\end{eqnarray}
where
\begin{eqnarray}
\gamma^{(l)} = \frac{(q^{(2l+1)} +q^{-(2l+1)})}{Q},
\end{eqnarray}
$l=0,\frac{1}{2},1,....\infty$ and $m$ runs by integer steps over the
range $-l\leq m \leq l$. In the following the states
$|v_{3}^{(l,m)},\gamma^{(l)}\rangle$ describing the boost will be notted
$|l,m\rangle$. They form an orthonormal basis
\begin{eqnarray}
\langle l,m|l',m'\rangle= \delta_{l',l}\delta_{m',m}
\end{eqnarray}
satisfying
\begin{eqnarray}
\gamma|l,m\rangle= \gamma^{(l)}|l,m\rangle~~and~~
V_{3}|l,m\rangle=v_{3}^{(l,m)}|l,m\rangle.
\end{eqnarray}
By setting $\Lambda_{0}^{~3}=\gamma V^{3}$, we obtain from (28)
$V^{3}=q-\frac{q}{\gamma^{2}(1+qV_{3})}$ and
$\Lambda_{3}^{~3}=q\gamma V_{3} +\frac{1}{\gamma(1+qV_{3})}$ from which we
deduce
\begin{eqnarray}
V^{3}|l,m\rangle=-q(\frac{q^{-2m}}{\gamma^{(l)}}-1)|l,m\rangle=v^{3(l,m)}
|l,m\rangle~~and~~
\Lambda_{3}^{~3}|l,m\rangle=(q^{2m}+q^{-2m}-\gamma^{(l)})|l,m\rangle.
\end{eqnarray}
From the orthogonality condition $G^{NM}\Lambda_{N}^{~0}\Lambda_{M}^{~0}=
G^{00}=-q^{-\frac{3}{2}}=-q^{-\frac{3}{2}}\gamma^{2} +q^{\frac{1}{2}}
\gamma^{2}(\frac{qV_{z}V_{\overline{z}}+q^{-1}V_{\overline{z}}V_{z}}{Q}+
V_{3}V_{3})$ and the commutation relation $V_{z}V_{\overline{z}}-
V_{\overline{z}}V_{z}=(q-q^{-1})QV_{3}(V_{3}+q^{-1})$ obtained from (17), we get
\begin{eqnarray}
V_{\overline{z}}V_{z}=\frac{1}{\gamma^{2}}(q^{-2}(1+qV_{3})
(1-q^{3}V_{3})-q^{-2}),\\
V_{z}V_{\overline{z}}=\frac{1}{\gamma^{2}}(q^{-2}(1+qV_{3})
(1-q^{-1}V_{3})-q^{-2}).
\end{eqnarray}
On the other hand, the commutation relations
\begin{eqnarray}
V_{3}V_{z}-q^{2}V_{z}V_{3}=(q-q^{-1})V_{z}~~and~~
V_{3}V_{\overline{z}}-q^{-2}V_{\overline{z}}V_{3}=-q^{-2}
(q-q^{-1})V_{\overline{z}}
\end{eqnarray}
obtained from (16), show that [3]
\begin{eqnarray}
V_{z}|l,m\rangle=(\alpha^{1}_{(l,m)})^{\frac{1}{2}}
|l,m+1\rangle~~and~~
V_{\overline{z}}|l,m\rangle=
(\alpha^{2}_{(l,m)})^{\frac{1}{2}}|l,m-1\rangle
\end{eqnarray}
where
\begin{eqnarray}
\alpha^{1}_{(l,m)}= \langle l,m|V_{\overline{z}}V_{z}|l,m\rangle=
\frac{1}{(\gamma^{(l)})^{2}}(q^{2m-1}Q\gamma^{(l)}-q^{4m}-q^{-2}),\\
\alpha^{2}_{(l,m)}= \langle l,m|V_{z}V_{\overline{z}}|l,m\rangle=
\frac{1}{(\gamma^{(l)})^{2}}(q^{2m-3}Q\gamma^{(l)}-q^{4m-4}-q^{-2}).
\end{eqnarray}
From (27) we get
\begin{eqnarray}
V^{z}=\frac{2}{Q\gamma}V_{z}(q^{-1}+V_{3})^{-1}~~and~~
V^{\overline{z}}=\frac{2q^{2}}{Q\gamma}V_{\overline{z}}(q^{-1}+V_{3})^{-1}
\end{eqnarray}
implying
\begin{eqnarray}
V^{z}|l,m\rangle=\frac{2}{Q}q^{-2m+1}V_{z}|l,m\rangle=\frac{2}{Q}q^{-2m+1}
(\alpha^{1}_{(l,m)})^{\frac{1}{2}}|l,m+1\rangle=
(\beta^{1}_{(l,m)})^{\frac{1}{2}}|l,m+1\rangle
\end{eqnarray}
where
\begin{eqnarray}
\beta^{1}_{(l,m)}= \langle l,m|V^{\overline{z}}V^{z}|l,m\rangle=
\frac{4}{(Q\gamma^{(l)})^{2}}(q^{-2m+1}Q\gamma^{(l)}-q^{-4m}-q^{+2}).
\end{eqnarray}
The same procedure applied to $V^{\overline{z}}$ gives
\begin{eqnarray}
V^{\overline{z}}|l,m\rangle=
(\beta^{2}_{(l,m)})^{\frac{1}{2}}|l,m-1\rangle
\end{eqnarray}
where
\begin{eqnarray}
\beta^{2}_{(l,m)}= \langle l,m|V^{z}V^{\overline{z}}|l,m\rangle=
\frac{4}{(Q\gamma^{(l)})^{2}}(q^{-2m+3}Q\gamma^{(l)}-q^{-4m-4}-q^{+2}).
\end{eqnarray}
Now we want to consider successive boost transformations given in terms of
left-coaction on the coordinates as:
\begin{eqnarray}
X''_{N}=(\Delta\otimes I)\Delta_{L}(X_{N})=(i\otimes\Delta_{L})
\Delta_{L}(X_{N})=\Lambda_{N}^{~K}\otimes\Lambda_{K}^{~M}\otimes X_{N}.
\end{eqnarray}
As notted above $X''_{N}$ fulfil the same commutation relation as $X_{N}$ and
$\Lambda_{~N}^{''M} =\Lambda_{N}^{~K}\otimes \Lambda_{K}^{~M}$ fulfil the
same commutation relations as $\Lambda_{N}^{~M}$. Then a state describing
the evolution of a particle in the coordinate system $X''_{N}$ read
$|{\cal P}''\rangle =|t'',x''_{3},\tau^{2}\rangle$. It is a common
eingenstate of $X''_{0}$, $X''_{3}$ and $\tau^{2}$ with eingenvalues $t''$,
$x''_{3}$ and $\tau^{2}$ respectively. As assumed above, the transformed
states may be written as
\begin{eqnarray}
|{\cal P}''\rangle=|sym_{q}''\rangle\otimes|{\cal P}\rangle=
|sym_{q}'\rangle\otimes|sym_{q}\rangle\otimes|{\cal P}\rangle=
|sym_{q}'\rangle\otimes|{\cal P}'\rangle
\end{eqnarray}
where $|sym_{q}''\rangle=|v''_{3},\gamma''\rangle$ is a common eigenstate of
$\gamma ''=\Lambda_{~0}^{''0}$ and $V''_{3}$ deduced from $\Lambda_{~3}^{''0} =
V''_{3}\gamma''$. Since $\Lambda_{~N}^{''M}$ fulfil the same commutation
relations as $\Lambda_{N}^{~M}$, $|v''_{3},\gamma''\rangle$ has the same form
as (37) and may be written as $|l_{3},m_{3}\rangle$
satifying
\begin{eqnarray}
\gamma''|l_{3},m_{3}\rangle=
\gamma^{(l_{3})}|l_{3},m_{3}\rangle~~and~~
V''_{3}|l_{3},m_{3}\rangle=
v_{3}^{(l_{3},m_{3})}|l_{3},m_{3}\rangle
\end{eqnarray}
where
\begin{eqnarray}
\gamma^{(l_{3})}=\frac{q^{(2l_{3}+1)}+q^{-(2l_{3}+1)}}{Q}~~,~~
v_{3}^{(l_{3},m_{3})}=q^{-1}(\frac{q^{2m_{3}}}{\gamma^{(l_{3})}} - 1)
\end{eqnarray}
and $l_{3}=0,\frac{1}{2},1,....\infty$ and $m_{3}$ runs by integer steps over the
range $-l_{3}\leq m_{3} \leq l_{3}$.\\
We are now ready to state the addition rule of the velocity out of the
coaction on the generators of the boost. In fact let
$|sym_{q}'\rangle =|l_{2},m_{2}\rangle$ and $|sym_{q}\rangle =
|l_{1},m_{1}\rangle$. For $l_{2}$ and $l_{1}$ fixed, the basis
$|l_{2},m_{2}\rangle\otimes|l_{1},m_{1}\rangle=
|l_{2},l_{1},m_{2},m_{1}\rangle$ contains $(2l_{2}+1)(2l_{1}+1)$ linear
independent states satisfying
\begin{eqnarray}
\langle l_{2},l_{1},m_{2},m_{1}|l_{2},l_{1},m'_{2},m'_{1}\rangle =
\delta_{m_{2},m'_{2}}\delta_{m_{1},m'_{1}}\\
\sum_{m_{2},m_{1}}|l_{2},l_{1},m_{2},m_{1}\rangle
\langle l_{2},l_{1},m_{2},m_{1}|=1.
\end{eqnarray}
The $m_{2}$ sum is performed over all values $-l_{2}\leq m_{2} \leq l_{2}$
and analogously in the $m_{1}$ case over the interval
$-l_{1}\leq m_{1}\leq l_{1}$. Thus in the basis
$|l_{2},l_{1},m_{2},m_{1}\rangle$ the state $|l_{3},m_{3}\rangle$ reads
\begin{eqnarray}
|l_{3},m_{3}\rangle =\sum_{m_{2},m_{1}} |l_{2},l_{1},m_{2},m_{1}\rangle
\langle l_{2},l_{1},m_{2},m_{1}|l_{3},m_{3}\rangle
\end{eqnarray}
where the coefficients $\langle l_{2},l_{1},m_{2},m_{1}|l_{3},m_{3}\rangle$
are the quantum-analog of the Clebsch-Gordon coefficients. Now we are ready
to state that
\begin{eqnarray}
m_{3}=m_{2}+m_{1}~~,~~~-l_{3}\leq m_{3} \leq l_{3}~~ and ~~~
|l_{2}-l_{1}|\leq l_{3} \leq l_{2}+l_{1}.
\end{eqnarray}
From the coaction on the boost generators we get
\begin{eqnarray}
\Delta(\Lambda_{0}^{~0})=\gamma ''= \Lambda_{0}^{~0}\otimes\Lambda_{0}^{~0} +
\frac{1}{2}(\Lambda_{0}^{~z}\otimes\Lambda_{\overline{z}}^{~0}+
\Lambda_{0}^{~\overline{z}}\otimes\Lambda_{z}^{~0})+
\Lambda_{0}^{~3}\otimes\Lambda_{3}^{~0}\\
\Delta(\Lambda_{3}^{~0})=V''_{3}\gamma ''=
\Lambda_{3}^{~0}\otimes\Lambda_{0}^{~0} +
\frac{1}{2}(\Lambda_{3}^{~z}\otimes\Lambda_{\overline{z}}^{~0}+
\Lambda_{3}^{~\overline{z}}\otimes\Lambda_{z}^{~0})+
\Lambda_{3}^{~3}\otimes\Lambda_{3}^{~0}.
\end{eqnarray}
By replacing (25) into (62) we get
\begin{eqnarray*}
V''_{3}\gamma ''=\Lambda_{3}^{~0}\otimes\Lambda_{0}^{~0} 
-\frac{q^{-1}}{2}(\Lambda_{0}^{~z}\otimes\Lambda_{\overline{z}}^{~0}&+&
\Lambda_{0}^{~\overline{z}}\otimes\Lambda_{z}^{~0})+
(\Lambda_{0}^{~0}+q\Lambda_{3}^{~0}-q^{-1}\Lambda_{0}^{~3})\otimes
\Lambda_{3}^{~0}=\\
\Lambda_{3}^{~0}\otimes(\Lambda_{0}^{~0} + q\Lambda_{3}^{~0})&+&
q^{-1}\Lambda_{0}^{~0}\otimes(\Lambda_{0}^{~0}+q\Lambda_{3}^{~0})\\
-q^{-1}(\Lambda_{0}^{~0}\otimes\Lambda_{0}^{~0}+
\frac{1}{2}(\Lambda_{0}^{~z}\otimes\Lambda_{\overline{z}}^{~0}&+&
\Lambda_{0}^{~\overline{z}}\otimes\Lambda_{z}^{~0})+\Lambda_{0}^{~3}\otimes
\Lambda_{3}^{~0})=\\
\Lambda_{3}^{~0}\otimes(\Lambda_{0}^{~0} + q\Lambda_{3}^{~0})&+&
q^{-1}\Lambda_{0}^{~0}\otimes(\Lambda_{0}^{~0}+q\Lambda_{3}^{~0})-
q^{-1}\gamma ''.
\end{eqnarray*}
By applying the latter relation to the state (59), we get
\begin{eqnarray}
V''_{3}\gamma ''|l_{3},m_{3}\rangle&=&v_{3}^{(l_{3},m_{3})}
\gamma^{(l_{3})}|l_{3},m_{3}\rangle =
q^{-1}(q^{2m_{3}}-\gamma^{(l_{3})})|l_{3},m_{3}\rangle=\\
\sum_{m_{2},m_{1}}(v_{3}^{(l_{2},m_{2})}\gamma^{(l_{2})}(1&+&qv_{3}^{(l_{1},m_{1})})
\gamma^{(l_{1})}|l_{2},l_{1},m_{2},m_{1}\rangle\langle l_{2},l_{1},m_{2},m_{1}
|l_{3},m_{3}\rangle +\nonumber\\
(q^{-1}\gamma^{(l_{2})}(1+qv_{3}^{(l_{1},m_{1})})\gamma^{(l_{1})}&-&q^{-1}
\gamma^{(l_{3})})|l_{2},l_{1},m_{2},m_{1}\rangle\langle l_{2},l_{1},m_{2},m_{1}
|l_{3},m_{3}\rangle=\\
\sum_{m_{2},m_{1}}q^{-1}(q^{2(m_{2}+m_{1})}&-&\gamma^{(l_{3})})
|l_{2},l_{1},m_{2},m_{1}\rangle
\langle l_{2},l_{1},m_{2},m_{1}|l_{3},m_{3}\rangle.
\end{eqnarray}
By identifying (63) with (65) and by applying
$\langle l_{2},l_{1},m_{2},m_{1}|$ from the left we get because of linear
independence
\begin{eqnarray}
(q^{2m_{3}}-q^{2(m_{2}+m_{1})})
\langle l_{2},l_{1},m_{2},m_{1}|l_{3},m_{3}\rangle=0
\end{eqnarray}
which implies $m_{3}=m_{2}+m_{1}$ and
\begin{eqnarray}
\langle l_{2},l_{1},m_{2},m_{1}|l_{3},m_{3}\rangle=0~~~if~~
m_{2}+m_{1} \not= m_{3}.
\end{eqnarray}
By applying (61) to (59) and then $\langle l_{2},l_{1},m_{2},m_{1}|$ from the
left, we get
\begin{eqnarray}
\gamma^{(l_{3})}\langle l_{2},l_{1},m_{2},m_{1}|l_{3},m_{3}\rangle&=&\\
(\gamma^{(l_{1})}q^{-2m_{2}}&+&
\gamma^{(l_{2})}q^{2m_{1}}-q^{-2(m_{2}-m_{1})})
\langle l_{2},l_{1},m_{2},m_{1}|l_{3},m_{3}\rangle\nonumber\\
&+&\frac{1}{2}(\beta^{1}_{(l_{2},m_{2}-1)}\alpha^{2}_{(l_{1},m_{1}+1)})^{\frac{1}{2}}
\langle l_{2},l_{1},m_{2}-1,m_{1}+1|l_{3},m_{3}\rangle \nonumber\\
&+&\frac{1}{2}(\beta^{2}_{(l_{2},m_{2}+1)}\alpha^{1}_{(l_{1},m_{1}-1)})^{\frac{1}{2}}
\langle l_{2},l_{1},m_{2}+1,m_{1}-1|l_{3},m_{3}\rangle.
\end{eqnarray}
For $m_{2}=l_{2}$ and $m_{1}=l_{1}$ or $m_{2}=-l_{2}$ and $m_{1}=-l_{1}$ the
second and third terms of the right hand side of this relation vanish
implying $l_{3}=l_{2}+l_{1}$. we note in these cases that the upper value
$m_{3}^{max}=l_{2}+l_{1}$ and the lower value $m_{3}^{min}=-(l_{2}+l_{1})$
appear once, thus the upper value of $l_{3}$ is $l_{3}^{max}=l_{2}+l_{1}$ and
\begin{eqnarray}
|l_{3}^{max},l_{3}^{max}\rangle=|l_{2},l_{1},l_{2},l_{1}\rangle,\\
|l_{3}^{max},-l_{3}^{max}\rangle=|l_{2},l_{1},-l_{2},-l_{1}\rangle.
\end{eqnarray}
Since they are $(2l_{2}+1)(2l_{1}+1)$ linear independent states
$|l_{2},l_{1},m_{2},m_{1}\rangle$ and $m_{3}^{max}=l_{2}+l_{1}$ appears only
once we may make a similar demonstration to the one of the addition of two
angular momentum, called triangle rule in the text book of quantum
machanics [7], to show that all values of
$l_{3}=l_{2}+l_{1},l_{2}+l_{1}-1,...|l_{2}-l_{1}|$ appear precisely once and
the number of states $|l_{3},m_{3}\rangle$ is equal the number of basis
$|l_{2},l_{1},m_{2},m_{1}\rangle$
\begin{eqnarray}
\sum_{|l_{2}-l_{1}|}^{l_{2}+l_{1}}(2l_{3}+1)=(2l_{2}+1)(2l_{1}+1).
\end{eqnarray}
Therefore, the projection of the velocity on the quantization direction is given in
terms of quantum number $m_{3}=m_{2}+m_{1}$ and
$l_{3}=l_{2}+l_{1},l_{2}+l_{1}-1,...|l_{2}-l_{1}|$ with
$-l_{3}\leq m_{3} \leq l_{3}$.\\
Now we are ready to compute explicitly the $q$-deformed Clebsch-Gordon
coefficients. We start from
\begin{eqnarray*}
\Delta(\Lambda_{z}^{~0})=\gamma ''V_{z}''&=&
\Lambda_{z}^{~0}\otimes\Lambda_{0}^{~0} +
\frac{1}{2}(\Lambda_{z}^{~z}\otimes\Lambda_{\overline{z}}^{~0}+
\Lambda_{z}^{~\overline{z}}\otimes\Lambda_{z}^{~0})+
\Lambda_{z}^{~3}\otimes\Lambda_{3}^{~0}=\\
&=&\Lambda_{z}^{~0}\otimes\Lambda_{0}^{~0} +
1\otimes\Lambda_{z}^{~0}+\Lambda_{z}^{~3}\otimes\Lambda_{3}^{~0}=\\
&=&\Lambda_{z}^{~0}\otimes(\Lambda_{0}^{~0} +q\Lambda_{3}^{~0})
+1\otimes\Lambda_{z}^{~0}
\end{eqnarray*}
where we have used (25), $\Lambda_{z}^{~z}=0$ and
$\Lambda_{z}^{~\overline{z}}=2$. By applying the latter relation to (59),
we obtain
\begin{eqnarray*}
\gamma^{(l_{3})}(\alpha^{1}_{(l_{3},m_{3})})^{\frac{1}{2}}|l_{3},m_{3}\rangle=
\sum_{m_{2},m_{1}}\gamma^{(l_{2})}(\alpha^{1}_{(l_{2},m_{2})})q^{2m_{1}}
|l_{2},l_{1},m_{2}+1,m_{1}\rangle
\langle l_{2},l_{1},m_{2},m_{1}|l_{3},m_{3}\rangle+\\
\gamma^{(l_{1})}(\alpha^{1}_{(l_{1},m_{1})})
|l_{2},l_{1},m_{2},m_{1}+1\rangle
\langle l_{2},l_{1},m_{2},m_{1}|l_{3},m_{3}\rangle
\end{eqnarray*}
which give because of linear independence the condition
\begin{eqnarray*}
\gamma^{(l_{3})}(\alpha^{1}_{(l_{3},m_{3})})^{\frac{1}{2}}
\langle l_{2},l_{1},m_{2},m_{1}|l_{3},m_{3}+1\rangle&=&\\
\gamma^{(l_{2})}(\alpha^{1}_{(l_{2},m_{2}-1)})^{\frac{1}{2}}q^{2m_{1}}
\langle l_{2},l_{1},m_{2}-1,m_{1}|l_{3},m_{3}\rangle &+&
\gamma^{(l_{1})}(\alpha^{1}_{(l_{1},m_{1}-1)})^{\frac{1}{2}}
\langle l_{2},l_{1},m_{2},m_{1}-1|l_{3},m_{3}\rangle.
\end{eqnarray*}
The same procedure gives for $\Delta(\Lambda_{\overline{z}}^{~0})=
\gamma''V_{\overline{z}}''$ the condition
\begin{eqnarray*}
\gamma^{(l_{3})}(\alpha^{2}_{(l_{3},m_{3})})^{\frac{1}{2}}
\langle l_{2},l_{1},m_{2},m_{1}|l_{3},m_{3}-1\rangle&=&\\
\gamma^{(l_{2})}(\alpha^{2}_{(l_{2},m_{2}+1)})^{\frac{1}{2}}q^{2m_{1}}
\langle l_{2},l_{1},m_{2}+1,m_{1}|l_{3},m_{3}\rangle &+&
\gamma^{(l_{1})}(\alpha^{2}_{(l_{1},m_{1}+1)})^{\frac{1}{2}}
\langle l_{2},l_{1},m_{2},m_{1}+1|l_{3},m_{3}\rangle
\end{eqnarray*}
These conditions give recursion relations for the calculation of
Clebsch-Gordon coefficients. For example in the case where
$l_{2}=\frac{1}{2}$ and $l_{1}=\frac{1}{2}$ we obtain the folowing
Clebsch-Gordon coefficients:
\begin{eqnarray}
\langle \frac{1}{2},\frac{1}{2},\frac{1}{2},-\frac{1}{2}|1,0\rangle=
(qQ)^{-\frac{1}{2}}~~and~~
\langle \frac{1}{2},\frac{1}{2},-\frac{1}{2},\frac{1}{2}|1,0\rangle=
(q^{-1}Q)^{-\frac{1}{2}}
\end{eqnarray}
leading to
\begin{eqnarray}
|1,1\rangle=|\frac{1}{2},\frac{1}{2},\frac{1}{2},\frac{1}{2}\rangle~~,~~
|1,-1\rangle=|\frac{1}{2},\frac{1}{2},-\frac{1}{2},-\frac{1}{2}\rangle,\\
|1,0\rangle=(qQ)^{-\frac{1}{2}}
|\frac{1}{2},\frac{1}{2},\frac{1}{2},-\frac{1}{2}\rangle+
(q^{-1}Q)^{-\frac{1}{2}}
|\frac{1}{2},\frac{1}{2},-\frac{1}{2},\frac{1}{2}\rangle,\\
|0,0\rangle=-(q^{-1}Q)^{-\frac{1}{2}}
|\frac{1}{2},\frac{1}{2},\frac{1}{2},-\frac{1}{2}\rangle+
(qQ)^{-\frac{1}{2}}
|\frac{1}{2},\frac{1}{2},-\frac{1}{2},\frac{1}{2}\rangle.
\end{eqnarray}
For the case $l_{2}=\frac{1}{2}$ and $l_{1}=1$ we obtain
\begin{eqnarray}
|\frac{3}{2},\frac{3}{2}\rangle=|\frac{1}{2},1,\frac{1}{2},1\rangle~~,~~
|\frac{3}{2},-\frac{3}{2}\rangle=|\frac{1}{2},1,-\frac{1}{2},-1\rangle,\\
|\frac{3}{2},\frac{1}{2}\rangle=(\frac{q^{-1}Q}{Q^{2}-1})^{\frac{1}{2}}
|\frac{1}{2},1,\frac{1}{2},0\rangle+q(\frac{1}{Q^{2}-1})^{\frac{1}{2}}
|\frac{1}{2},1,-\frac{1}{2},0\rangle ,\\
|\frac{3}{2},-\frac{1}{2}\rangle=(\frac{qQ}{Q^{2}-1})^{\frac{1}{2}}
|\frac{1}{2},1,-\frac{1}{2},0\rangle+q^{-1}(\frac{1}{Q^{2}-1})^{\frac{1}{2}}
|\frac{1}{2},1,\frac{1}{2},-1\rangle\\
|\frac{1}{2},\frac{1}{2}\rangle=q(\frac{1}{Q^{2}-1})^{\frac{1}{2}}
|\frac{1}{2},1,\frac{1}{2},0\rangle-(\frac{q^{-1}Q}{Q^{2}-1})^{\frac{1}{2}}
|\frac{1}{2},1,-\frac{1}{2},0\rangle ,\\
|\frac{1}{2},-\frac{1}{2}\rangle=-q^{-1}(\frac{1}{Q^{2}-1})^{\frac{1}{2}}
|\frac{1}{2},1,-\frac{1}{2},0\rangle+(\frac{qQ}{Q^{2}-1})^{\frac{1}{2}}
|\frac{1}{2},1,\frac{1}{2},-1\rangle.
\end{eqnarray} \\
\\
{\bf Conclusion:}\\
In this paper we have showed how the addition rule of the velocity in the
noncommutative special relativity is derived from the left coaction (53) on a
quantum coordinate system of the noncommutative Minkowski space-time. The
states describing the evolution of a particle in different coordinate systems
tied by quantum boost cotransformations (54) are computed explicitely and the
recursion formulas giving the $q$-deformed Glebsch-Gordon coefficients of
the transformed states are investigated.\\
\\
{\bf Acknowledgments:} I am particularly grateful to the Abdus Salem
International Center for Theoretical Physics, Trieste, Italy.\\

{\bf References:}\\
1)U. Meyer, Commun. Math. Phys. 174(1996)457. P. Podle\'s, Commun. Math. Phys.
181(1996)569.\\
2)M. Pillin, W. B. Schmidke and J. Wess, Nucl. Phys. B(1993)223. M. Pillin
and Weilk, J. Phys. A: Math. Gen. 27(1994)5525. B. L. Cerchiai and J. Wess
:"q-Deformed Minkowski Space based on q-Lorentz Algebra" LMU-TPW 98-02,
MPI-PhT/89-09. O. Ogiesvestsky, W. B. Schmidke, J. Wess and B. Zumino, Commun.
Math. Phys. 150(1992)495.\\
3)M. Lagraa, "On the noncommutative special relativity", Math.ph/9904014.\\
4)P. Podles and S. L. Woronowicz, Commun. Math. Phys. 130(1990)381.\\
5)M. Lagraa, J. Geom. Phys 34(2000)206.\\
6)S. L. Woronowicz, Publ. RIMS, Kyoto Univ. 23 (1987) 117, Commun. Math.
Phys. 122(1989)125. P. Podle\'s, Lett. Math. Phys. 14(1987)193, Lett. Math.
Phys. 18(1989)107, Commun. Math. Phys. 170(1995)1.\\
7)W. Greiner and Berndt Muller,"Quantum Mechanics, Symmetries" second revised
edition, Spinger-Verlag (1994). A. Messiah, "Mecanique Quantique", tome 2,
Dunod, Paris 1960.
\end{document}